# Fabrication and Characterization of InAs/AlSb based Magnetic Hall Sensors

Marsha Mary Parmar, Hurmal Saren and Pintu Das
*Department of Physics*
*Indian Institute of Technology Delhi*
Hauzkhas, New Delhi-110016, India
Marsha.Mary.Parmar@ee.iitd.ac.in

*Abstract*—Hall effect based magnetic field sensors are vector sensors which are sensitive to the perpendicular component of the magnetic field. Typically, Si-based Hall sensors are used for most usages requiring low field sensitivity. For higher sensitivity, III-V semiconductors such as GaAs, InSb, etc., provide better alternatives due to higher carrier mobilities at room temperature. In this work, Hall sensors based on two-dimensional electron gas (2DEG) at the heterostructure of InAs/AlSb are fabricated. The 2DEG is about 20 nm below the surface. Carrier density of $3.2 \times 10^{16}$ /m$^2$ and mobility of 1.8 m$^2$/Vs were deduced via Hall measurements at RT. 1/f-noise analyses were carried out to calculate the magnetic field resolution of ~1 µT/Hz$^{1/2}$.

## I. INTRODUCTION

Hall effect sensors are by far the most widely used magnetic sensors with wide range of applications such as home appliances (speed control in vacuum cleaner, drum speed control in washers, etc[1].), automotive technology (steering angle sensing, stroke sensing, etc[2].), and robotics (proximity and contact sensor)[3]. On the other hand, the Hall effect has also been used in fundamental research to detect very minute changes in magnetic flux to study the physics of superconducting vortex melting[4], domain wall motion in magnetic thin film samples[5], etc. For these measurements very high sensitivity of Hall devices is required. Hall devices made of semiconductors harvest the galvanometric effect on charge carriers, transforming the magnetic signal into an electric one. While the most celebrated magnetic field sensors are silicon-based, interest has been grown in group III-V devices such as GaAs (0.1-0.8 m$^2$/Vs) [6],[7], InSb (2.5 m$^2$/Vs) [8], [9], etc., due to high mobility ranges compared to silicon-based devices. GaAs-AlGaAs heterostructure based Hall sensors exhibit ultra-high sensitivity at low temperatures (T<100 K), which has been used for studies of domain wall pinning and depinning at Peierl's potential [5], [10] quasi-static switching behavior in artificial spin ice systems with defects [11], etc. Another very interesting magnetic sensors are superconducting quantum interference devices (SQUID) which can detect magnetic field ranging from tens of femto-tesla to 9 T [12]. While the aforementioned magnetic sensors exhibit extreme sensitivity levels, the major drawback with these sensors is the temperature levels at which they operate (T<100 K). The sensitivity of Hall sensors scales with the electron mobility of the semiconductor material [13]. Hence, for application purpose sensors with high electron mobility at room temperature are required. In this sense, two-dimensional electron gas (2DEG) presents an excellent choice in terms of high electron mobility (1-4 m$^2$/Vs) [13], [14]. A 2DEG is formed when electrons are confined in an interface between two different materials, such as heterostructures of group III-V semiconductors (AlInSb/InSb, InAs/AlSb). In this work, we report the fabrication and characterization of Hall devices based on the heterojunction of InAs/AlSb for room temperature applications. While the transport properties of electrons in 2DEG is material dependent, it is observed that the fabrication process also affects the electronic transport in 2DEG [15]. Hence, the fabrication process of InAs/AlSb Hall bars is discussed in detail. Further, to estimate the magnetic field levels that need to be detected using the Hall effect sensors fabricated in this work, we simulate the disturbances caused by a ferromagnetic object due to their own remnant magnetization. These disturbances cause distortions in an otherwise uniform geomagnetic field (50 µT). We simulate the absolute change in Earth's magnetic field as a function of distance from the center of a ferromagnetic sphere with a relative magnetic permeability of 1000.

## II. DEVICE FABRICATION

### A. InAs/AlSb heterostructure

AlSb which is the spacer layer, has a bandgap of 1.6 eV and InAs which is the channel layer, has a bandgap of 0.4 eV. Hence, at the interface of AlSb and InAs, a quantum well is formed with a large conduction band offset of 1.35 eV shown in Fig. 1 [18], [21]. The quantum well thus formed restricts the motion of electrons classically in the z-direction, forming a 2DEG. InAs/AlSb heterostructure was grown on a semi-insulating (SI) GaAs substrate. During the growth process, Arsenic (As) antisite defects were deliberately introduced as it pins the Fermi level to the center of the band gap. This facilitates the substrate with a minimum concentration of electrons and holes, which is desirable. To reduce the lattice mismatch between GaAs and InAs, buffer layers were grown. The buffer layers not only provide a smooth surface for 2DEG but also suppress the impurities that are inevitably present in and on the substrate and tend to migrate towards the epitaxial layer during the growth process. The channel layer (InAs) is sandwiched between two spacer layers (AlSb), has a thickness of 12.5 nm. To further increase the electron concentration, a second spacer layer of AlSb was grown with Te doping 3 nm away from InAs layer. Finally, capping layers are grown to protect the AlSb layer against oxidation. Fig. 2 shows the schematic of the layer sequence for InAs/AlSb heterostructure used in this work. The heterostructure was grown by molecular beam epitaxy (MBE) at Intelligent Epitaxy Technology. Inc. USA.

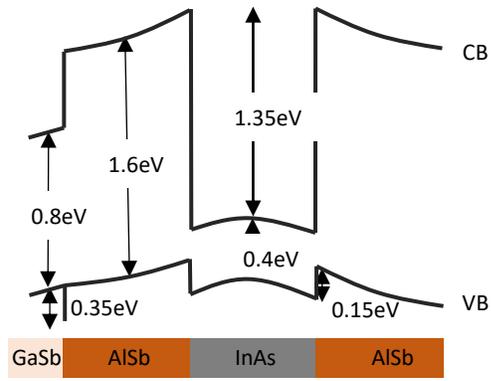

Fig. 1 Schematics of InAs/AlSb band alignment for the heterostructure. CB and VB represent conduction and valence band edges.

*B. Fabrication of Hall bars*

The wafer containing InAs/AlSb heterostructure was cut into $5 \times 5$ mm$^2$ for device fabrication. The small chips were cleaned in acetone, followed by isopropyl alcohol (IPA) and deionized (DI) water. The chips were transferred into an ultrasonication bath for the removal of surface contamination. It is crucial to reduce the intensity of the ultrasonication bath for InAs/AlSb chips. We found that prolonged exposure to sonication bath results in damaging of chips. After sonication, samples were carefully dried with N$_2$ gas.

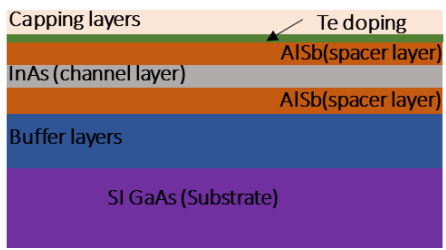

Fig. 2 Schematics of InAs/AlSb heterostructure used for magnetic Hall sensors.

The Hall bar pattern was transferred to the sample via photolithography system (SF-100 Xpress, make: ScoTech, UK). The size of the Hall bars fabricated in this work is of $20 \times 20$ μm$^2$. For the lithography process, photoresist S1805 (Shipley) and developer M26A (Shipley) was used. M26A is an ammonia-based developer, which slowly etches group III-V semiconductor [16]. Therefore, it is essential to process development after ensuring optimal exposure. Next, wet etching was used to create Hall bar pattern of the 2DEG which lies 23 nm from the surface for our samples [17]. The chip was dipped into a solution of HCl:H$_2$O$_2$:H$_2$O :: 1:1:50 for 30 sec followed by HCl dip of 2 sec. With this, an etch depth of about 35 nm was achieved from the top of the surface. As shown in Fig. 3a an array of Hall bars are fabricated with each hall bar of $20 \times 20$ μm$^2$ size. Atomic microscopy (AFM) was used to confirm the etch-depth as shown in Fig. 3b. A second layer of lithography was done to selectively open the contact pad area ($250 \times 250$ μm$^2$) for metal deposition. Before deposition, we etched the top layers again using the same recipe mentioned above to expose InAs channel layer. The samples were immediately taken to e-beam deposition chamber for deposition of 5 nm of Cr and 70 nm of Au. After deposition, samples were immersed in acetone for the lift-off process. We used wedge bonding (wire bonder make: TPT, Germany) and optimized the bonding parameter using gold wire of diameter 25 μm. Finally, the chips were placed in a 20-pin lead less chip carrier (LCC) for electrical characterization. The final device after wire bonding is shown in Fig. 4.

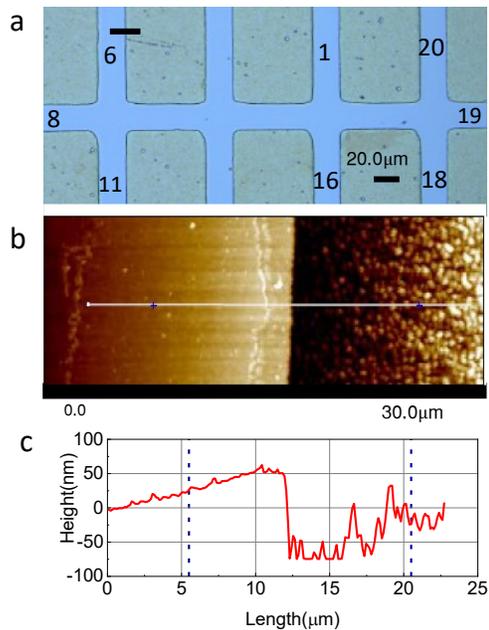

Fig. 3 Etch depth analysis of Hall bar array: (a) Optical image of the etched Hall bar (20 μm × 20 μm) array. Numbers are to identify the contacts for measurement purpose. (b) AFM image of a part of Hall device, shown by dark rectangle in contact 6 in (a) after etching. (c) Profile of the etched part along the line shown in (b).

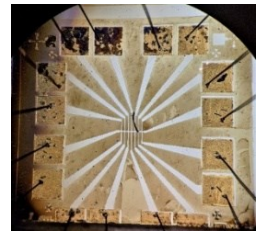

Fig. 4 Optical image of the final device, which is wire bonded on a 20-pin LCC chip carrier.

## III. ELECTRICAL CHARACTERIZATION

### A. Room Temperature (RT) Electrical Characterization:

To check the quality of contacts with 2DEG, four-probe DC measurements were performed. Linear I-V data confirm the ohmic nature of contacts as shown in Fig. 5 at RT. The contact resistance of ~20 kΩ was calculated for our devices. We also determine the room temperature resistivity of $\rho_\square = 111.10\ \Omega$.

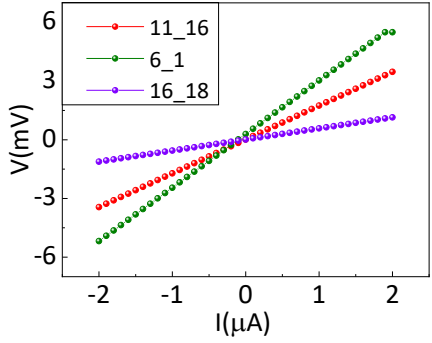

Fig. 5 Four-probe dc I-V measurements across different contacts of the devices. Labels indicate the exact contact nos for the device (see Fig.3(a)).

### B. Hall measurements

Hall measurements were performed at room temperature within the magnetic field range of ± 1.0 T, applied perpendicular to the device. Fig. 6a show the schematic of Hall measurements for our devices. As shown in Fig 6a dc current of 1μA is sent through the current probe while the Hall voltage is measured across probe 1-3. The Hall voltage across probe3 is shown in Fig. 6b as a function of the magnetic field. The negative slope of the measured Hall voltage clarifies the majority charge carriers as electrons. An offset of 40 µV is observed in the Hall voltage which is due to the misalignment of Hall bars during the fabrication process, as shown in 6a. Hall coefficient, $R_H (= V_H/IB)$ of 194.4 Ω/T and the corresponding electron density of $n_D = 3.21 \times 10^{16}/m^2$ were determined from the Hall voltage data. From the I-V measurements shown in Fig. 5, Hall mobility of $\mu = 1.8\ m^2/Vs$ is calculated for RT. Electron mean free path $l_e = \frac{\hbar\mu}{e}\sqrt{\frac{n}{2\pi}} \approx 0.53$ µm clearly indicates that the transport measurements are performed in the diffusive regime [18].

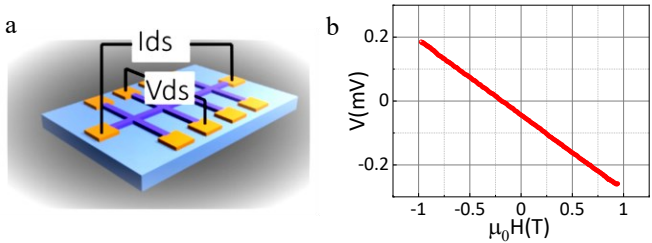

Fig. 6 (a) Schematic of Hall measurement set-up. (b) Hall measurements at room temperature (RT). The offset from zero crossing is due to the asymmetry in the Hall device fabrication.

## IV. 1/F NOISE MEASUREMENTS

Field sensitivity of 2DEG based Hall sensors critically depends on the inherent electronic noise of the system. The most crucial is the 1/f-noise type, which limits the sensitivity of the Hall devices in the low-frequency regime. To determine the magnetic field sensitivity of our device we performed ac noise spectroscopic measurements in five probe geometry [19]. In Fig. 7 we show the power spectral density (PSD), $S_v$ of voltage noise for I = 0 and 10 µA, respectively in log-log scale. For I = 0 µA, $S_v$ is independent of frequency and shows the floor (background) noise of the experimental set-up. The peaks at 25 Hz and 70 Hz are due to extrinsic noise and do not relate to the sample. $S_v$ for I = 10 µA shows the $1/f^\alpha$-type behavior (α = 1.2). A deviation from $1/f^\alpha$-type behavior is observed for f > 30 Hz which is not clear at present. In the following, noise values have been taken by fitting the data for frequency up to 10 Hz.

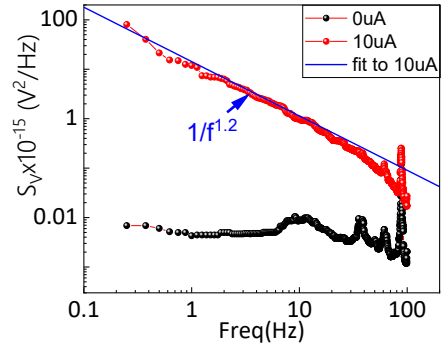

Fig. 7 Power spectral density (Sv) of voltage noise for the InAs/AlSb heterojunction for I = 0 µA and 10 µA, respectively. $1/f^\alpha$ nature of noise, with $\alpha$ values between 1 - 1.2 is observed. The black curve measured at 0 $\mu A$ current, gives the experimental noise background. Peaks within the range of 10-100 Hz are due to extrinsic noise from the background.

To ensure the measured noise is intrinsic to the sample, we performed current dependence of $S_v$. Fig. 8 shows that $S_v \propto I^2$ confirming the intrinsic nature of $1/f^\alpha$- noise at 1 Hz from the linear fit of the spectra shown in Fig. 8.

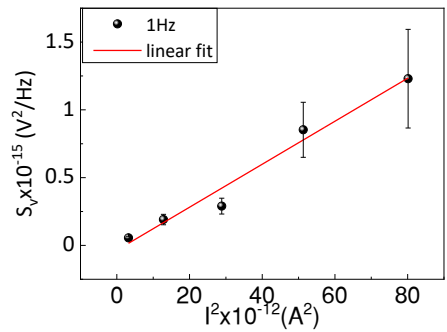

Fig. 8 Linear variation of voltage power spectral density at 1 Hz with $I^2$.

Assuming $S_v$ is the noise voltage PSD for the Hall device, we determine the minimum field resolution $B_{min}$ using $\sqrt{S_v}/R_H I$ to be of the order of 1 µT at 100 Hz bandwidth [18].

## V. Summary

In summary, we have successfully shown the fabrication and characterization of magnetic Hall sensors at the heterojunction of InAs/AlSb heterostructure. Noise spectroscopy measurements on the Hall device were performed, and a magnetic field resolution of 1µT was deduced for the fabricated magnetic field sensors. The Hall sensor fabricated in this work can detect the magnetic signature of ferromagnetic object.

## Acknowledgment

PD acknowledges the financial support from MHRD and DRDO through IMPRINT project "SILDS." We acknowledge the fabrication and characterization facilities at Nano Research Facility, IITD.